\documentclass[final,5p,times,twocolumn]{elsarticle}
\usepackage{amsmath,latexsym,amssymb,geometry}
\usepackage{setspace,stackrel,tikz,graphicx,caption}
\usepackage{exscale,relsize,stackrel}
\usepackage{MnSymbol}

\newcommand{\EE}{\mathrm{E}}
\newcommand{\BMw}{\mbox{\boldmath $w$}}
\newcommand{\Iota}{\mathrm{i}}
\newcommand{\BMx}{\mbox{\boldmath $x$}}

\begin{document}

\begin{frontmatter}	
	\title{Adaptive Blind Sparse-Channel Equalization}
	\author[Habib]{Shafayat~Abrar}
	\address[Habib]{Associate Professor\\ School of Science and Engineering\\ Habib University, Gulistan-e-Jauhar\\ Block 18, Karachi 75290, Pakistan\\ Email: \textrm{\texttt{shafayat.abrar@sse.habib.edu.pk}}}

\begin{abstract}
In this article, a fractional-norm constrained blind adaptive algorithm is presented for sparse channel equalization. In essence, the algorithm improves on the minimisation of the constant modulus (CM) criteria by adding a sparsity inducing \(\ell_p\)-norm penalty. Simulation results demonstrate that the proposed regularised equalizer exploits the inherent channel sparsity effectively and exhibits faster convergence compared to its counterparts.     	

\end{abstract}
\begin{keyword}
	Blind equalization; constant modulus algorithm; sparse channel; adaptive filter; channel equalization
\end{keyword}
\end{frontmatter}


\section{Introduction}

The constant modulus algorithm (CMA) is a widely studied and, under certain conditions, an admissible solution for adaptive blind channel equalization problem \cite{JohnsonProcIEEE}. The performance of the traditional
CMA, however, is not satisfactory if the underlying channel is \textit{sparse}. By sparse channel, it is meant that the number of significant channel coefficients is
much less than its total dimensionality.
To make CMA suitable for such channels, Matrin \textit{et al.} \cite{Martin} devised a number of sparse versions of CMA where they incorporated sparsity under the \textit{approximate natural gradient} (ANG) framework and developed \textit{proportionate-type} updates (i.e. the changes in the equalizer parameters were proportional to their magnitudes). These variants happened to perform better than CMA on sparse channels but exhibited significant jitter when forced to converge faster.
More recently, \textit{regularised} sparse solutions have attracted serious attention in adaptive signal processing community. Most of these efforts have been centered around the sparsity promoting minimisation of \(\ell_0\) and \(\ell_1\) norms of the filter parameters \cite{gu2009norm,shi2011adaptive,angelosante2010online,shi2010convergence}. The use of \textit{fractional-norm} regularisation has also evolved as an admissible candidate and has been found sparser than \(\ell_1\) and more tractable computationally than \(\ell_0\) \cite{xu2012regularization,wu2013gradient}.

In this work, motivated by the idea of norm-constrained optimisation \cite{Douglas2000gradient}, we design a sparse CMA by projecting the gradient vector of the cost onto an \(\ell_p\)-ball and exploiting the smallest geometrical angle between the gradient vectors associated with the cost and the constraint. We discuss the stability of the proposed update, and provide simulation results to demonstrate its superiority over the CMA and sparse variants of CMA.

\section{Proposed Algorithm}

Consider the following instantaneous constant modulus (CM) cost function subjected to a constraint for sparsity:
\begin{equation}\label{EqCostConst}
\min_{\boldsymbol{w}} J(\boldsymbol{w}_k)={\textstyle\frac{1}{2}}\!\left(R - \boldsymbol{w}_k^H\boldsymbol{x}_k\boldsymbol{x}_k^H\boldsymbol{w}_k\right)^2,
~\mathrm{s.t.}~\|\boldsymbol{w}_k\|_p^p\le c
\end{equation}
where \(\boldsymbol{w}_k=[w_{1,k},w_{2,k},\cdots,w_{N,k}]^T\) is an $N\times 1$ linear finite-impulse response equalizer vector, \(\boldsymbol{x}_k=[x_k,x_{k-1},\cdots,x_{k-N+1}]^T\) is an $N\times 1$ channel observation vector, $R>0$ is a statistical constant \cite{JohnsonProcIEEE}, and \(\|\boldsymbol{w}_k\|_p\) is a pseudo $\ell_0$-norm defined as \(\|\boldsymbol{w}_k\|_p:= (\sum_{i=1}^N|w_{i,k}|^p)^{1/p}\).
The objective is to mitigate the sparse channel interference and recover the transmitted sinal using solely the equalizer output $\boldsymbol{w}_k^H\boldsymbol{x}_k$. The a priori information about the sparsity of channel is assumed to be available, and as a result, the equalizer coefficients are also assumed to be parameterised with sparse representation. By sparse,we mean that the number of significant parameters in $\boldsymbol{w}_k$, $M$, is
much less than its total dimensionality (that is \(M\ll N\)).

\begin{figure}[ht!]\centering
\includegraphics[scale=0.5, bb=70 167 518 604, clip]{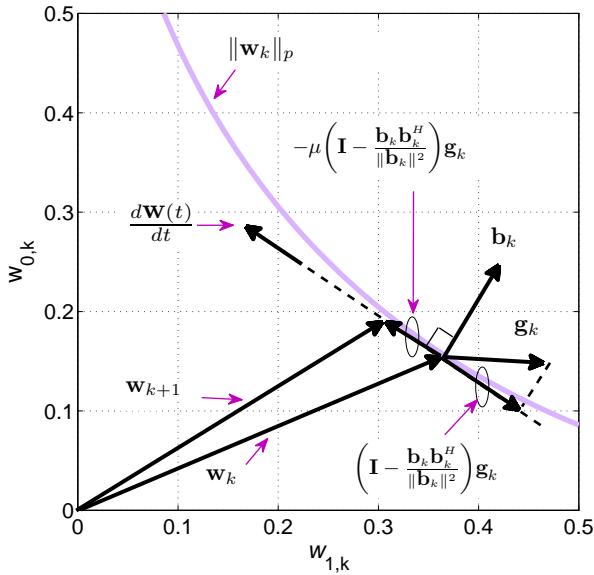}
\caption{Geometrical interpretation of constrained optimization.}
\label{figuregeometry}
\end{figure}

Note that for the feasible set \(Q:=\|\boldsymbol{w}_k\|_p^p\le c\), the minimum value in CM cost is assumed to be attainable
as the objective is continuous (and admissible for equalizable channels), and the set $Q$ is compact. Also the set $Q$ is nonconvex, so there might be multiple projection points in general on the geodesic of \(\|\boldsymbol{w}_k\|_p^p= c\). For
the purpose of equalization, however, any such minimiser is acceptable. The Lagrangian for (\ref{EqCostConst}) reads
\begin{equation}
L(\boldsymbol{w}_k,\lambda_k)= (R - \boldsymbol{w}_k^H\boldsymbol{x}_k\boldsymbol{x}_k^H\boldsymbol{w}_k )^2
-\lambda_k (\|\boldsymbol{w}_k\|_p^p-c),
\end{equation}
where $\lambda_k$ is a real-valued Lagrangian multiplier. The gradient-based update for the minimisation is obtained as
\begin{equation}
\boldsymbol{w}_{k+1}=\boldsymbol{w}_k-\mu\, \partial L(\boldsymbol{w}_k,\lambda_k)\big/\partial \boldsymbol{w}_k^\ast,
\end{equation}
where the superscript $\ast$ denotes complex conjugate. Denoting \(\boldsymbol{g}_k:=\partial J(\boldsymbol{w}_k)\big/\partial \boldsymbol{w}_k^\ast\) and \(\boldsymbol{b}_k:=\partial \|\boldsymbol{w}_k\|_p^p\big/\partial \boldsymbol{w}_k^\ast\) as two gradient vectors, we get
\(\boldsymbol{w}_{k+1}=\boldsymbol{w}_k-\mu (\boldsymbol{g}_k-\lambda_k\boldsymbol{b}_k )\).
We have to select $\lambda_k$ such that
\(\|\boldsymbol{w}_{k+1}\|_p=c,~\forall k\),
i.e., the $\ell_p$-norm of vector $\boldsymbol{w}_k$ is conserved for all values of $k$. This property yields a flow equation in continuous time-domain:
\begin{equation}\label{EqIsonormal}
\frac{d\|\boldsymbol{w}(t)\|_p^p}{dt}=\left(\frac{\partial \|\boldsymbol{w}(t)\|_p^p}{\partial \boldsymbol{w}(t)^\ast}\right)^{\!H}\,\frac{d\boldsymbol{w}(t)}{dt}=:\boldsymbol{b}(t)^H\frac{d\boldsymbol{w}(t)}{dt}=0
\end{equation}
where the superscript $H$ denotes the complex conjugate
transpose operation. The two vectors $\boldsymbol{b}(t)$  and $d\boldsymbol{w}(t)\big/dt$ are orthogonal to each other, and they are normal and tangential to the surface \(\|\boldsymbol{w}(t)\|_p^p=c\) at $\boldsymbol{w}(t)$, respectively.
Moreover, for a sufficiently small $\delta$, we can approximate the time derivative as follows:
\begin{equation}\label{EqLambda}
\left.\frac{d\boldsymbol{w}(t)}{dt}\right|_{t=k\delta}\approx\lim_{\delta\rightarrow 0}\frac{\boldsymbol{w}_{k+1}-\boldsymbol{w}_k}{\delta}=-\mu\frac{(\boldsymbol{g}_k-\lambda_k\boldsymbol{b}_k)}{\delta}
\end{equation}
Combining (\ref{EqIsonormal}) and (\ref{EqLambda}), we obtain an optimal value of $\lambda_k$, as given by
\begin{equation}\label{EqLambdaValue}
\boldsymbol{b}_k^H\left(\boldsymbol{g}_k-\lambda_k\boldsymbol{b}_k\right)=0~~~\Rightarrow~~~\lambda_k=\dfrac{\boldsymbol{b}_k^H\boldsymbol{g}_k}{\|\boldsymbol{b}_k\|^2}
\end{equation}
 The vector \(\lambda_k\boldsymbol{b}_k={\boldsymbol{b}_k^H\boldsymbol{g}_k\boldsymbol{b}_k}\big/{\|\boldsymbol{b}_k\|^2}\) is the component of $\boldsymbol{g}_k$ projected onto $\boldsymbol{b}_k$. The weight update computes the projection of $\boldsymbol{g}_k$ onto $\boldsymbol{b}_k^{\bot}$ which is given by \(\boldsymbol{g}_k-\lambda_k\boldsymbol{b}_k=\big(\mathbf{I}-\boldsymbol{b}_k\boldsymbol{b}_k^H/\|\boldsymbol{b}_k\|^2\big)\boldsymbol{g}_k\). So that the required update is not only against the gradient $\boldsymbol{g}_k$ but also follows the geodesic of $\|\boldsymbol{w}_k\|_p^p$. Refer to Fig.~\ref{figuregeometry} for the geometrical illustration for a real-valued two-tap equalizer. Moreover, the term $\lambda_k\boldsymbol{b}_k$ serves as \textit{zero-point attraction} \cite{jin2010stochastic}, because it reduces the distance between $\boldsymbol{w}_k$ and the origin when $|\boldsymbol{w}_k|$ is small.

Note that $\lambda_k$ is (a sort of) complex-valued cosine \cite{Scharnhorst2001angles} of the angle between $\boldsymbol{b}_k$ and $\boldsymbol{g}_k$. From the problem definition, however, the value of $\lambda_k$ is required to be real-valued.
To obtain a real-valued $\lambda_k$, we have a lemma from the theory of holomorphic geometry of complex vectors \cite[\textit{Lemma 2.2.2}]{Goldman1999complex} \big(below \(\langle\boldsymbol{a},\boldsymbol{b}\rangle=\boldsymbol{a}^H\boldsymbol{b}\), and $\Re$ denotes the real part\big):

\textit{Lemma 1:
Let $\langle\cdot,\cdot\rangle$ be a positive Hermitian form on a complex vector space $E_{\mathbb{C}}$. The underlying
real vector space $E_{\mathbb{M}}$ inherits a positive definite inner product. Let \(\boldsymbol{v}_1,\boldsymbol{v}_2\in E_{\mathbb{C}}\) be non-zero vectors. They span complex lines \(\mathbb{C}\boldsymbol{v}_i\subset E_{\mathbb{M}}\) whose angle satisfies \[\angle(\mathbb{C}\boldsymbol{v}_1,\mathbb{C}\boldsymbol{v}_2)\le \angle(\boldsymbol{v}_1, \boldsymbol{v}_2).\] The algebraic expression for the \underline{smallest} angle, independent of spanning, in terms of the Hermitian structure is obtained from
\[\cos\big(\angle(\mathbb{C}\boldsymbol{v}_1,\mathbb{C}\boldsymbol{v}_2)\big)={\big|\langle \boldsymbol{v}_1,\boldsymbol{v}_2\rangle\big|}\big/\big({\|\boldsymbol{v}_1\|\,\|\boldsymbol{v}_2\|}\big).\]
}

\noindent\textit{Proof: Replace $\boldsymbol{v}_i$ by $\boldsymbol{v}_i/\|\boldsymbol{v}_i\|$ to assume that \(\boldsymbol{v}_1,\boldsymbol{v}_2\) are unit vectors. Minimising
the angle between them, in $\mathbb{C}\boldsymbol{v}_1$ and $\mathbb{C}\boldsymbol{v}_2$, is equivalent to maximising its cosine.
The real-valued cosine between unit vectors in these planes equals
 \[\cos\big(\angle(e^{\mathbf{i}\psi_1}\boldsymbol{v}_1, e^{\mathbf{i}\psi_1}\boldsymbol{v}_2)\big)=\Re\big[e^{\mathbf{i}(\psi_1-\psi_2)}{\langle\boldsymbol{v}_1,\boldsymbol{v}_2 \rangle}\big]/\big({\|\boldsymbol{v}_1\|\,\|\boldsymbol{v}_2\|}\big),\] for \(\psi_1, \psi_2\in \mathbb{M}\). The maximum value of this expression over all \(\psi_1, \psi_2\in \mathbb{M}\)
equals $|\langle\boldsymbol{v}_1,\boldsymbol{v}_2 \rangle|$ as desired.
}

Owing to \textit{Lemma 1}, the optimal value of $\lambda_k$ is obtained as
\begin{equation}\label{EqSolutions}
\lambda_{k,\textrm{optimal}}=\dfrac{\big|\boldsymbol{b}_k^H\boldsymbol{g}_k\big|}{\|\boldsymbol{b}_k\|^2} \end{equation}
Since \(\lambda_{k,\textrm{optimal}}>0\), therefore the resulting algorithm maximises the $\ell_p$-ball of the equalizer coefficients until it coincides with the extremum of minimising CM cost.

\begin{figure}[ht!]\centering
\includegraphics[scale=0.5, bb=84 250  527 564, clip]{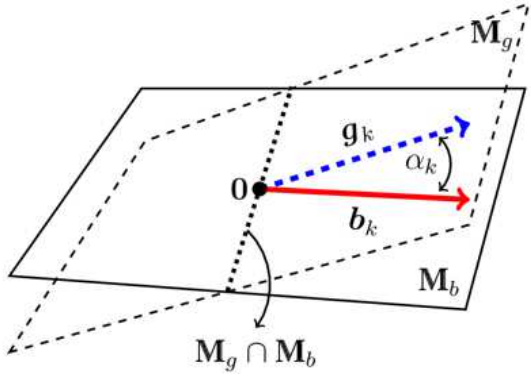}
\caption{The geometry of the planes spanned by $\boldsymbol{b}_k$ and $\boldsymbol{g}_k$, and the minimum angle between them.}\label{FigureGeometry2}
\end{figure}

\noindent \textbf{Remark}: Let \(\mathbb{M}_{g}:=\mathrm{span}(\{\boldsymbol{g}_k\})\) and \(\mathbb{M}_{b}:=\mathrm{span}(\{\boldsymbol{b}_k\})\) be two complex-valued $N$-dimensional vector (sub)spaces. The orthogonal projection theorem suggests that the minimum angle $\alpha_k(\mathbb{M}_{g},\mathbb{M}_{b})$ (or the maximum cosine $c(\mathbb{M}_{g},\mathbb{M}_{b})$) between $\mathbb{M}_g$ and $\mathbb{M}_b$ is defined by (see Fig.~\ref{FigureGeometry2})
\begin{equation}\label{EqPropGeo54}\begin{aligned}
c(\mathbb{M}_{g},\mathbb{M}_{b}):=\sup%
\bigg\{\big|\langle\boldsymbol{g}_k,\boldsymbol{b}_k\rangle\big|:%
\boldsymbol{g}_k\in\mathbb{M}_{g}\cap(\mathbb{M}_{g}\cap\mathbb{M}_b)^\bot, &\\ \boldsymbol{b}_k\in\mathbb{M}_{b}\cap(\mathbb{M}_{g}\cap\mathbb{M}_b)^\bot,\|\boldsymbol{g}_k\|^2=\|\boldsymbol{b}_k\|^2=1\bigg\}&
\end{aligned}\end{equation}

The complex-valued cosine of the angle between two complex vectors $\boldsymbol{v}_1$ and $\boldsymbol{v}_2$ is given generally as \(\cos(\theta_C):=\cos\big(\angle(\boldsymbol{v}_1, \boldsymbol{v}_2)\big)={\langle \boldsymbol{v}_1,\boldsymbol{v}_2\rangle}\big/\big({\|\boldsymbol{v}_1\|\,\|\boldsymbol{v}_2\|}\big)\) \(=:\rho e^{\mathbf{i}\theta_K}\), where $\theta_C\in\mathbb{C}$ is called the complex angle, and \(\rho:=|\cos(\theta_C)|=:\cos(\theta_H)\le 1\). The angles \(0\le \theta_H\le \pi/2\) and \(-\pi<\theta_K\le\pi\) are known as the \textit{Hermitian angle},
and the \textit{Kasner's pseudo-angle}, respectively, between the vectors $\boldsymbol{v}_1$
and $\boldsymbol{v}_2$. So the proposed equalizer $\ell_p$-SCMA exploits the Hermitian angle in the update process which is not only the smallest angle but is insensitive to the multiplication of the vectors by any complex scalars, and it is desirable in the context of CM equalization where the equalizer update is required to be insensitive to multiplication by complex exponentials which represents phase/frequency offset errors.

So we find an adaptive solution to the sparse equalization problem (\ref{EqCostConst}) which minimises the CM criterion, moves along the geodesic of the constraint surface, and exploits the smallest angle between the gradient vectors (possibly spanning complex lines), as given by:
\begin{equation}\label{EqPropAlgo}
\boldsymbol{w}_{k+1}=\boldsymbol{w}_k-\mu\left(\boldsymbol{g}_k-\dfrac{\big|\boldsymbol{b}_k^H\boldsymbol{g}_k\big|}{\|\boldsymbol{b}_k\|^2}\boldsymbol{b}_k\right),\\
\end{equation}
where $\boldsymbol{g}_k$ and $\boldsymbol{b}_k$ are specified as:
\begin{subequations}\label{EqPropAlgo2}\begin{alignat}{1}
&\boldsymbol{g}_k=\left( \boldsymbol{w}_k^H\boldsymbol{x}_k\boldsymbol{x}_k^H\boldsymbol{w}_k-R\right)\boldsymbol{x}_k\boldsymbol{x}_k^H\boldsymbol{w}_k,\\
&\boldsymbol{b}_k={\frac{p}{2}}\left[\frac{w_{1,k}}{|w_{1,k}|^{2-p}},\cdots\cdots,\frac{w_{N,k}}{|w_{N,k}|^{2-p}}\right]^T.
\end{alignat}\end{subequations}

\section{Steady-state Stability}

The update (\ref{EqPropAlgo}) is stable. Denoting \(\gamma=|\boldsymbol{b}_k^H\boldsymbol{g}_k|/\|\boldsymbol{b}_k\|^2\), we obtain energy of the update (\ref{EqPropAlgo}) as given by
\begin{equation}\begin{split}
\|\boldsymbol{w}_{k+1}\|^2&=\|\boldsymbol{w}_k\|^2+\mu^2\|\boldsymbol{g}_k\|^2
+\mu^2\gamma^2{\|\boldsymbol{b}_k\|^2}\\
&~~~~~~~~+2\mu\Re\big[\boldsymbol{w}_k^H\big(\gamma\boldsymbol{b}_k - \boldsymbol{g}_k\big) - \mu\gamma^2{\|\boldsymbol{b}_k\|^2}\big]\\
&=\|\boldsymbol{w}_k\|^2+\mu^2\|\boldsymbol{g}_k\|^2
-\mu^2\gamma^2{\|\boldsymbol{b}_k\|^2}+2\mu\Re\big[\boldsymbol{w}_k^H\big(\gamma\boldsymbol{b}_k - \boldsymbol{g}_k\big)\big].
\end{split}\end{equation}
Owing to Bussgang theorem \cite{BDHaykin1994}, we have
\(
\EE\boldsymbol{w}_k^H\big(\gamma\boldsymbol{b}_k - \boldsymbol{g}_k\big)=0.
\) Further
exploiting the independency between $\boldsymbol{x}_k$ and $\boldsymbol{w}_k$ (independence theorem \cite{Mazo1980}), we obtain \(\EE\big(\|\boldsymbol{g}_k\|^2
-\gamma^2{\|\boldsymbol{b}_k\|^2}\big)=0,\)
yielding \(\EE\|\boldsymbol{w}_{k+1}\|^2
=\EE\|\boldsymbol{w}_k\|^2\) which implies that
there is no growth in the energy of $\boldsymbol{w}_k$ and thus proves the stability.

\section{Explicit Regularization}

We may add a second stage to perform explicit regularization. The aim of this stage is to (further) prune the equalizer coefficients as obtained in the first stage by introducing explicit $\ell_{1/2}$ or $\ell_{2/3}$ regularization as a brute-force method to prevent the coefficients taking large values, and to drive the unnecessary coefficients (which fall below certain threshold) to zero. Elegant closed-form solutions for $\ell_{1/2}$ or $\ell_{2/3}$ regularization have been developed by Zhang and Ye \cite{Zhang2015l2}.

Consider the following lemma:\\

\noindent\textit{Lemma 2 \cite{Miao2015general}:
  Let $f$ denote the objective function \(\min_h f=(h-w)^2+\lambda |h|^p\), \(0<p<1\), \(\lambda>0\). It has a unique minimum $h_\ast$ for \(|w|\ge \tau(p,\lambda)\), where
  \begin{equation}\label{hbgjhgjh}
    \tau(p,\lambda)=\frac{2-p}{2}(1-p)^{\frac{p-1}{2-p}}\lambda^{\frac{1}{2-p}}.
  \end{equation}
}

Next we discuss closed form solutions for \(p=1/2\) and \(p=2/3\).

\subsection{Closed-form solution for \(\ell_{1/2}\) regularization}

Once we have $\BMw_k=\BMw_k^R+\Iota \BMw_k^I$ from the update (\ref{EqPropAlgo}), we need to regularize $\BMw_k^R$ and $\BMw_k^I$ separately as formulated below:
\begin{subequations}\label{EqSecondEqualizer}
\begin{alignat}{1}
&\boldsymbol{h}_{k}^R=\arg\min_{\boldsymbol{h}}\left\{\left(\boldsymbol{h}-\boldsymbol{w}_{k}^R\right)^2+\lambda_R\|\boldsymbol{h}\|_{1/2}^{1/2}\right\}\\
&\boldsymbol{h}_{k}^I=\arg\min_{\boldsymbol{h}}\left\{\left(\boldsymbol{h}-\boldsymbol{w}_{k}^I\right)^2+\lambda_I\|\boldsymbol{h}\|_{1/2}^{1/2}\right\}
\end{alignat}
\end{subequations}
where \(\boldsymbol{h}\in\mathbb{M}^N\) is an auxiliary variable. The closed form solution to the above optimization problems are given as (below $L$ denotes either $R$ or $I$):
\begin{subequations}\label{EqSecondEqualizer2}
\begin{alignat}{1}
&{h}_{k,i}^L=\left\{\begin{array}{rr}
\!\!\!\frac{2}{3}|w_{k,i}^L|\left(1+\cos\left(\frac{2\pi}{3}-\frac{2}{3}\phi(w_{k,i}^L)\right)\right), & \!\! w_{k,i}^L > \frac{\sqrt[3]{54}}{4}\lambda_L^{\frac{2}{3}} \\
\!\!\!-\frac{2}{3}|w_{k,i}^L|\left(1+\cos\left(\frac{2\pi}{3}-\frac{2}{3}\phi(w_{k,i}^L)\right)\right), & \!\! w_{k,i}^L < -\frac{\sqrt[3]{54}}{4}\lambda_L^{\frac{2}{3}} \\
0, & \textrm{otherwise}
                    \end{array}\right.\\
&\phi(w_{k,i}^L)=\arccos\left(\textstyle\frac{3\sqrt{3}\lambda_L}{8}\,
|w_{k,i}^L|^{-\frac{3}{2}}\right),~~~i=1,2,\cdots,N
\end{alignat}
\end{subequations}
The regularized equalizer second-stage output is thus obtained as \(s_k=\boldsymbol{h}_k^H\BMx_k\), where \(\boldsymbol{h}_k=\boldsymbol{h}_k^R+\Iota \boldsymbol{h}_k^I\).

\begin{figure}[ht!]\centering
\setlength{\unitlength}{.4in}
\begin{picture}(5,2.2)(-2,-0.3)
\linethickness{1pt}
\put(0,0){\line(1,0){4}}
\put(4,0){\line(-4,3){2}}
\put(0,0){\line(4,3){2}}
\put(2,-.25){\makebox(0,0){$y_1^2-1$}}
\put(3.1,1){\makebox(0,0){$y_1$}}
\put(1,1){\makebox(0,0){$1$}}
\put(2,1.7){\makebox(0,0){$C$}}
\put(4.2,0){\makebox(0,0){$A$}}
\put(-0.2,0){\makebox(0,0){$B$}}
\end{picture}
\caption{Triangular interpretation for solving cubic polynomial.}\label{FigTriangle}
\end{figure}
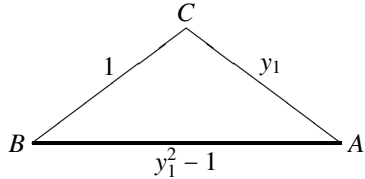

\noindent \textit{Proof:}
Consider a scalar $\ell_{1/2}$ optimization problem as follows:
\begin{equation}\label{Eqbasicprob1}
  \min_h \bigg\{(h-w)^2+\lambda|h|^{1/2}\bigg\}
\end{equation}
Taking derivative with respect to $h$, and substituting to zero, we get
\begin{equation}\label{Eqbasicprob554544}
h-w+\frac{\lambda}{4\sqrt{|h|}}\textrm{sign}(h)=0
\end{equation}
Substituting \(|h|=z^2\), we obtain
\begin{equation}
z^3\textrm{sign}(h)-wz+\frac{\lambda}{4}\textrm{sign}(h)=0
\end{equation}
Note that we require \(h<0\,(>0)\) for \(w<0\,(>0)\). Let \(h<0\), it gives \(\textrm{sign}(h)=-1\), and \(w=-|w|\), and we obtain \begin{equation}\label{Eqsimplemodew}z^3-|w|z+\frac{\lambda}{4}=0\end{equation} Same is the result when \(w>0\); so we proceed with (\ref{Eqsimplemodew}). In order to have three real-valued roots, we need to ensure that the Cardan's discriminant\footnote{For a cubic polynomial \(z^3+cz+d=0\), the Cardan's discriminant is defined as \(\delta:=-4c^3-27d^2\).} is \textit{positive}, this gives
\[
\frac{|w|^3}{3^3}-\frac{\lambda^2}{4^3}>0 ~~\Rightarrow~~|w|>\frac{3}{4}\lambda^{2/3}.
\]
Due to Lemma 2, however, we can find solution only if \(|w|>\frac{\sqrt[3]{54}}{4}\lambda^{2/3}>\frac{3}{4}\lambda^{2/3}\). Further substituting \(z=y\sqrt{{|w|}/{3}}\), we get
\begin{equation}\label{Eqbasicprob5454544}
y^3-3y-2q=0
\end{equation}
where
\(
q:=-{(\lambda}/{8})\left({3}/{|w|}\right)^{3/2}
\). Eq. (\ref{Eqbasicprob5454544}) may be solved by considering the triangle in Fig.~\ref{FigTriangle}, where $y_1$ represents one of the three roots of $y$. We outline the proof as conceived by Mitchell \cite{Mitchell200791} in the interest of readers.

 Using the cosine law, we obtain
\begin{equation}\label{Eqbasic6686882354}
-q=:\cos(C)=\frac{1+y_1^2-(y_1^2-1)^2}{2y_1}=\frac{y_1(3-y_1^2)}{2}
\end{equation}
which justifies the claim that $y_1$ is one of the roots of (\ref{Eqbasicprob5454544}). Similarly we obtain
\begin{equation}\label{Eqbasicjbhghghjghn}
\cos(B)=\frac{y_1^2-2}{2},~~\textrm{and}~~\cos(A)=\frac{y_1}{2}.
\end{equation}
From (\ref{Eqbasicjbhghghjghn}), we obtain \(B=2A\). Since \(A+B+C=\pi\), therefore \(A=\frac{1}{3}\pi-\frac{1}{3}C\). Now employing the sine law, we obtain
\begin{equation}\label{Eqbasicbgfngjhgvgvgng}
\frac{\sin(A)}{1}=\frac{\sin(B)}{y_1}=\frac{\sin(C)}{y_1^2-1}
\end{equation}
which implies \(y_1=\sin(B)/\sin(A)\), and gives
\begin{equation}\label{Eqbasicfbgjhgjhbmjhbnj}
z_1=\sqrt{\frac{|w|}{3}}\frac{\sin(A)}{\sin(B)}
=\sqrt{\frac{4|w|}{3}}\cos(A)
=\sqrt{\frac{4|w|}{3}}\cos\!\bigg(\frac{\pi}{3}-\frac{C}{3}\bigg)
\end{equation}
The other two ($z_2$ and $z_3$) roots may be found by adding $\pm\frac{2\pi}{3}$ in the argument of $\cos(\cdot)$; however, by inspecting these roots, we find that the root specified in (\ref{Eqbasicfbgjhgjhbmjhbnj}) is the desired root. From (\ref{Eqbasicfbgjhgjhbmjhbnj}), we obtain the desired value of $h$ as follows:
\begin{equation}\label{Eqbasicfbgjhgjhnj}
h=\left\{\begin{array}{l}
           \frac{2}{3}w\bigg(1+\cos\!\bigg(\frac{2}{3}\pi-\frac{2}{3}C\bigg)\bigg) \\
           \quad\quad\quad\quad\textrm{for}~|w|>\frac{\sqrt[3]{54}}{4}\lambda^{2/3} \\ \\
           0,\quad\quad\quad\textrm{otherwise}
         \end{array}\right.
\end{equation}

\subsection{Closed-form solution for \(\ell_{2/3}\) regularization}

Consider a scalar $\ell_{2/3}$ optimization problem as follows:
\begin{equation}\label{Eqbasicprob2}
  \min_h \bigg\{(h-w)^2+\lambda|h|^{2/3}\bigg\}
\end{equation}
The solution of above is rigourously presented in \cite{Cao2013fast,Zhang2015l2}. Here, we sketch a similar proof but in simpler steps.
Taking derivative with respect to $h$, and equating to zero, we get
\begin{equation}\label{Eqbasicprobjhgnjhhb}
h-w+\frac{\lambda}{3}\frac{\textrm{sign}(h)}{|h|^{1/3}}=0
\end{equation}
Substituting \(|h|=z^3\), we get
\begin{equation}\label{Eqbasicprob5454}
z^4\,\textrm{sign}(h)-wz+\frac{\lambda}{3}\,\textrm{sign}(h)=0
\end{equation}
We exploit Ferrari's idea to introduce a parameter $t$ in (\ref{Eqbasicprob5454})
\begin{equation}\label{Eqbasicprob54jhbgjhgnkhjhgjhg}
(z^2+t)^2=(2t)z^2+|w|z+\bigg(t^2-\frac{\lambda}{3}\bigg)
\end{equation}
such that the right hand side becomes a \textit{monic} quadratic polynomial in $z$, i.e., it has a real root with multiplicity $2$, or equivalently the discriminant is zero, which gives
\begin{equation}\label{Eqbasicprob54jhbgjhgnkhhgmjhgnhhg}
|w|^2-4(2t)\bigg(t^2-\frac{\lambda}{3}\bigg)=0~~\Rightarrow~~\bigg(t^2-\frac{\lambda}{3}\bigg)=\frac{|w|^2}{8t}
\end{equation}
Substituting the value of $t^2-\frac{\lambda}{3}$ in (\ref{Eqbasicprob54jhbgjhgnkhjhgjhg}), we obtain:
\begin{equation}\label{Eqbasicprob54jhbgjhgnkhj}
(z^2+t)^2=\bigg(\sqrt{2t}z+\frac{|w|}{\sqrt{8t}}\bigg)^2
\end{equation}
which gives
\begin{equation}\label{Eqbasicprob54jhbhgnkhj}
z^2+t=\pm\bigg(\sqrt{2t}z+\frac{|w|}{\sqrt{8t}}\bigg)
\end{equation}
Above the two roots associated with the negative sign are of no use, as they lead to an undesirable result $h<0$ ($h>0$) for $w>0$ ($w<0$). Solving, however, for positive sign, we obtain
\begin{equation}\label{Eqbasicprob54jhbhjhgbmjhghj}
z=\sqrt{\frac{t}{2}}\pm\sqrt{\frac{|w|}{\sqrt{8t}}-\frac{t}{2}}
\end{equation}
where the root of our interest is the one with plus sign as follows:
\begin{equation}\label{Eqbasicprob54jhbhjhgbghj}
z=\sqrt{\frac{t}{2}}+\sqrt{\frac{|w|}{\sqrt{8t}}-\frac{t}{2}}
\end{equation}
The last task is to find out the value of $t$ from the cubic expression (\ref{Eqbasicprob54jhbgjhgnkhhgmjhgnhhg}); we specify it again
\begin{equation}\label{Eqbasicprob54jhbkjbhnkjhjkjkgnhhg}
t^3-\frac{1}{3}\lambda\, t-\frac{1}{8}|w|^2=0
\end{equation}
Evaluating the Cardan's discriminant, we obtain
\begin{equation}\label{Eqbasicprob54jhbkjbjnhbjhnhjhg}
\begin{aligned}
  \delta& =-4\left(-\frac{\lambda}{3}\right)^3-27\left(-\frac{|w|^2}{8}\right)^2\\
  & = \frac{4}{27}\lambda^3 - \frac{27}{64}|w|^4\\
  & < \frac{4}{27}\lambda^3 - \frac{27}{64}\frac{16}{81}3\lambda^3~~ \left(|w|>\frac{2}{3}\sqrt[4]{3\lambda^3} ~\textrm{due to Lemma~2} \right)\\
  & < -\frac{11}{108}\lambda^3\\
  & <0~~~ \left(\because~\lambda >0\right)
\end{aligned}
\end{equation}
which implies that there is only one real-valued root of (\ref{Eqbasicprob54jhbkjbhnkjhjkjkgnhhg}). Since Mitchell's triangle method requires $\delta>0$, therefore it cannot help us find that root. Using Holmes formula \cite{Holmes200286}, however, we immediately obtain the required real-valued root of (\ref{Eqbasicprob54jhbkjbhnkjhjkjkgnhhg}) in a closed-form as follows:
\begin{equation}\label{Eqbasicprob5bjhgjhgjhgggnhhg}
t=\frac{2}{3}\sqrt{\lambda}\cosh\left(\frac{1}{3}\cosh^{-1}\!\left(\frac{27}{16}|w|^2\lambda^{-3/2}\right)\right).
\end{equation}
Owing to the relation \(|h|=z^3\), we obtain $h$ as follows:
\begin{equation}\label{Eqbashbjhgjhgjhgjhnj}
h=\left\{\begin{array}{l}
           \textrm{sign}(w)\left(\sqrt{\dfrac{t}{2}}+\sqrt{\dfrac{|w|}{\sqrt{8t}}-\dfrac{t}{2}}\right)^3 \\
           \quad\quad\quad\quad\textrm{for}~|w|>\frac{2}{3}\sqrt[4]{3\lambda^3} \\ \\
           0, ~~\textrm{otherwise}.
         \end{array}\right.
\end{equation}
where $t$ is as specified in (\ref{Eqbasicprob5bjhgjhgjhgggnhhg}).

\section{Simulation Results}
\begin{figure}[ht!]\centering
  \includegraphics[scale=0.45]{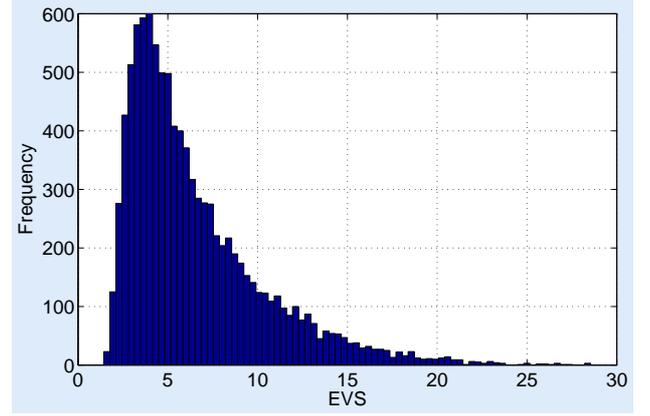}
\caption{EVS histogram obtained from ten thousand randomly generated sparse channels. The mean and standard deviation of the EVS are $6.57$ and $3.88$, respectively.}
\label{figureHisto}
\end{figure}

We compare the proposed two-stage regularized sparse CMA (RSCMA) equalizer with the traditional CMA \cite{JohnsonProcIEEE} and two sparse variants of CMA like ANG-CMA \cite{Martin} and SCMA($p$) \cite{khalid2015blind}.
The baseband model of the sparse channels have \(100\) taps with five non-zero taps obtained using the following program:
\begin{verbatim}
h=zeros(1,100);  i0=randi([1,10],1,1);
i1=randi([20,30],1,1); i2=randi([40,50],1,1);
i3=randi([70,80],1,1); i4=randi([90,100],1,1);
h(i0)=0.1*(2*rand-1)+0.1*(2*rand-1)*1i;
h(i1)=1+(2*rand-1)*1i;
h(i2)=0.5*(2*rand-1)+0.2*(2*rand-1)*1i;
h(i3)=0.2*(2*rand-1)+0.2*(2*rand-1)*1i;
h(i4)=0.1*(2*rand-1)+0.1*(2*rand-1)*1i;
h(tt,:)=h(tt,:)/norm(h(tt,:))';
\end{verbatim}
The average eigenvalue spread of the channels obtained from the above program (\texttt{sparse\_channel.m}) is nearly 4.8 with standard deviation 1.8. The histogram of the eigen-value spread is shown in Fig.~\ref{figureHisto}.

\begin{figure}[ht!]\centering
\begin{tabular}{c}
  \includegraphics[scale=0.35]{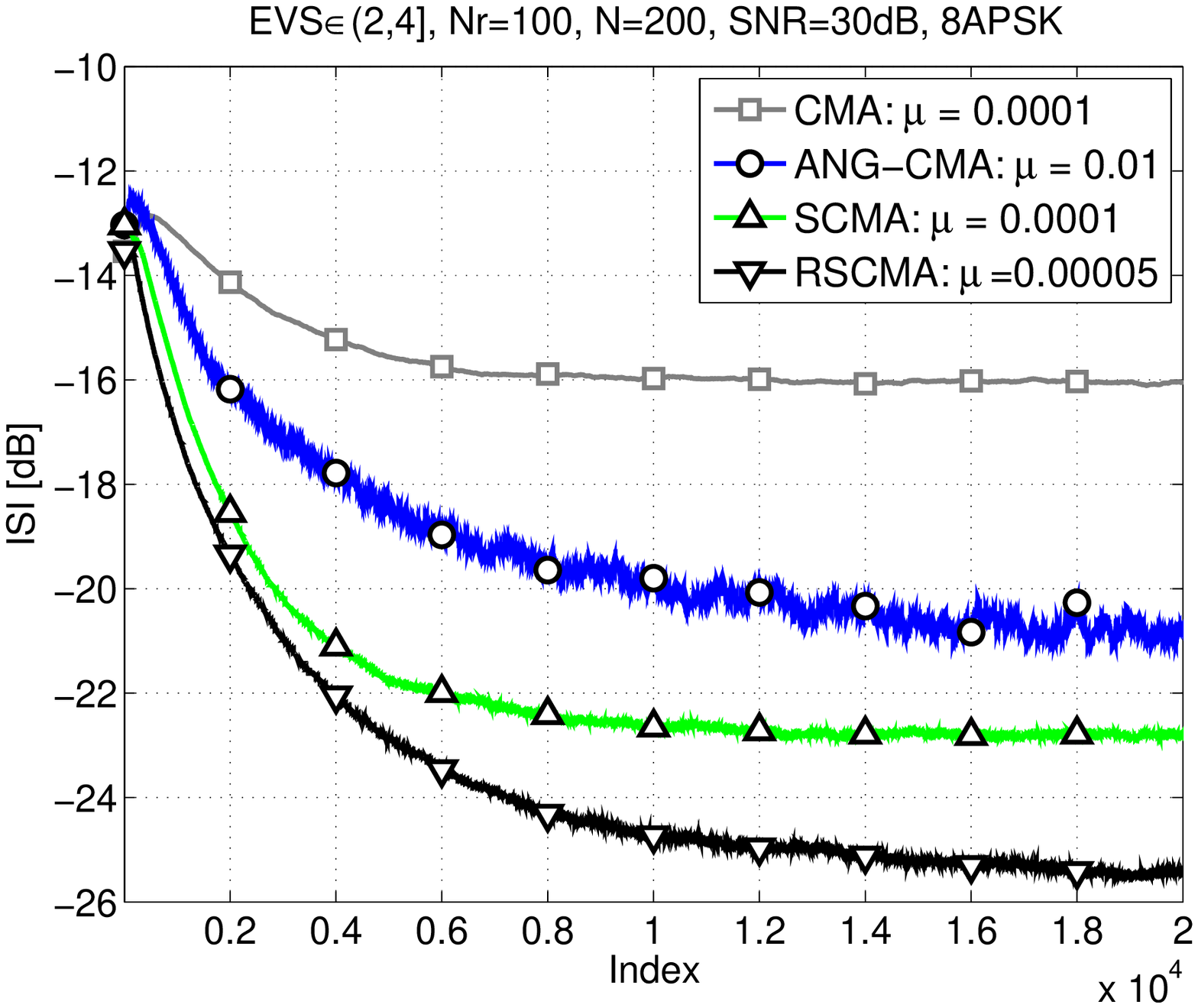} \\ \footnotesize{(a)}\\
  \includegraphics[scale=0.35]{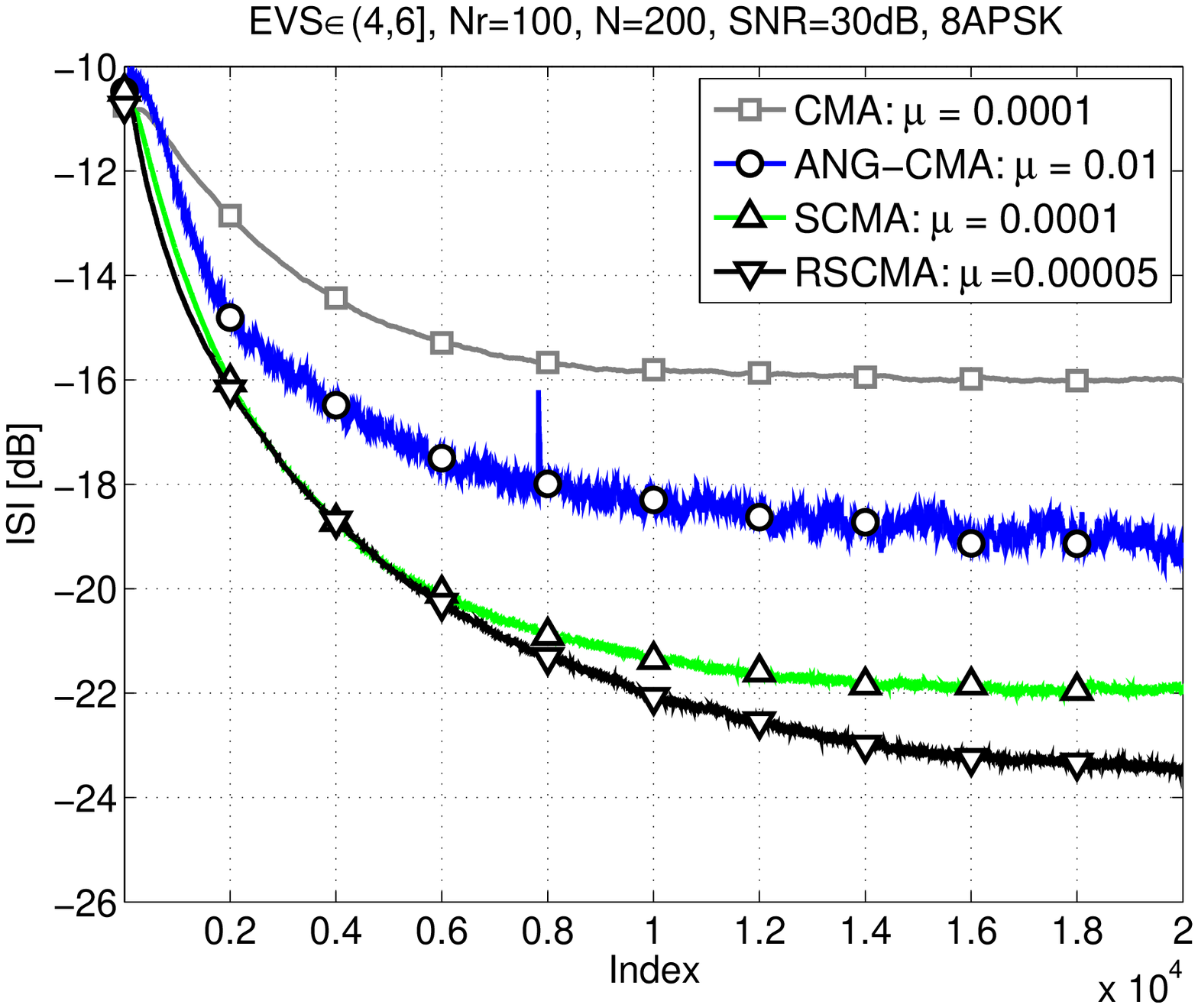} \\
  \footnotesize{(b)}\\
  \includegraphics[scale=0.35]{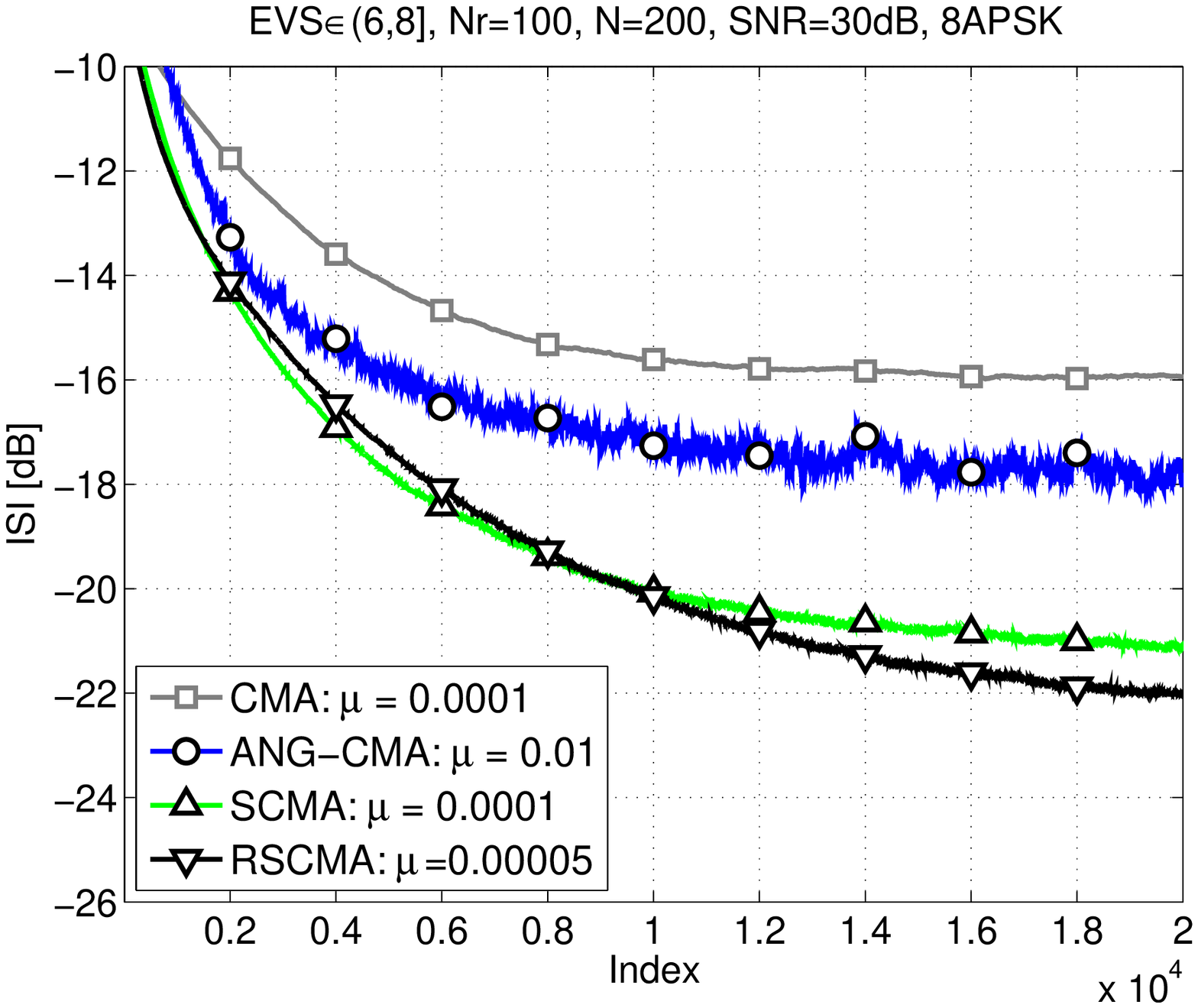} \\
  \footnotesize{(c)}\\
  \includegraphics[scale=0.35]{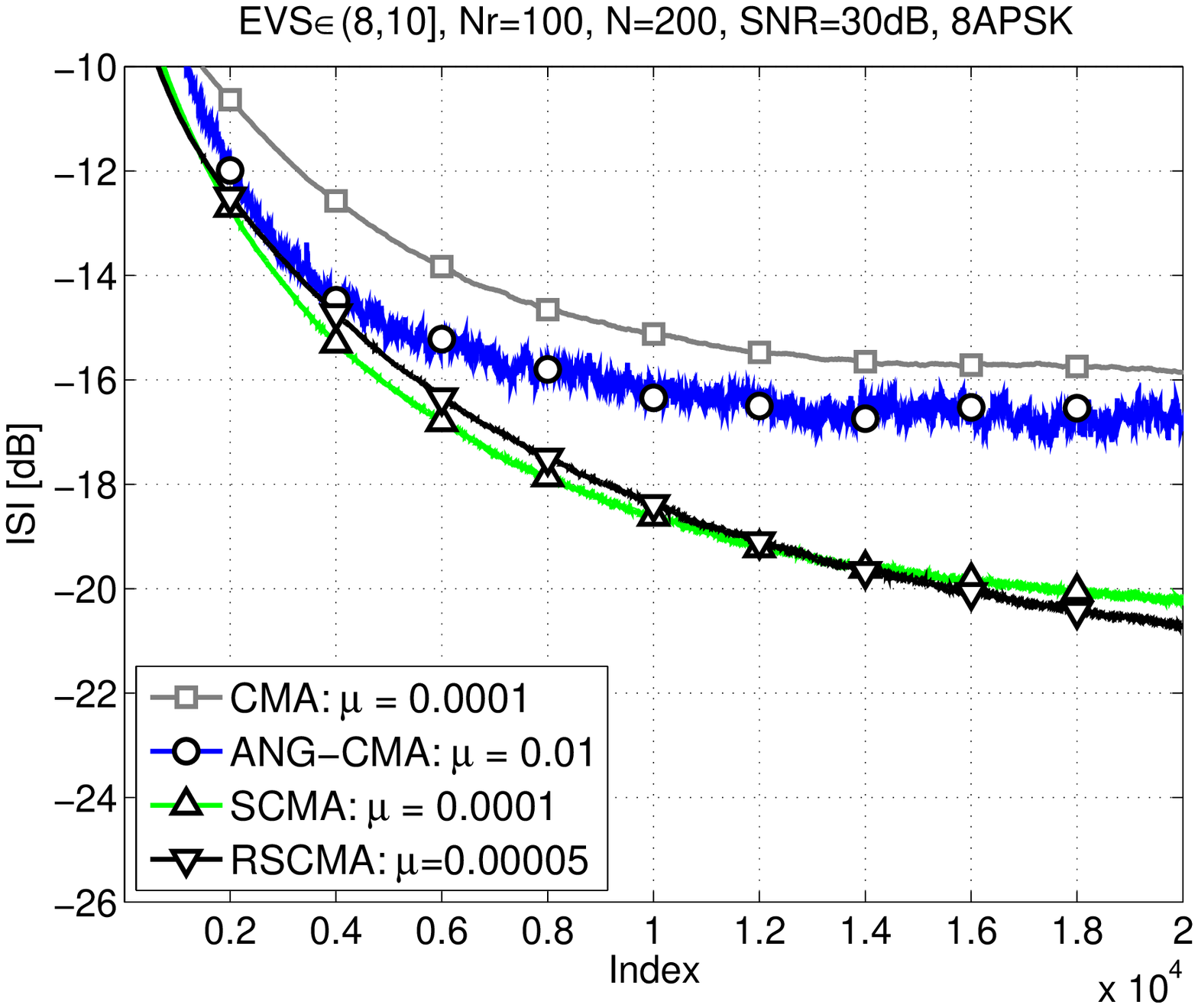} \\
  \footnotesize{(d)}
\end{tabular}
\caption{Comparison of residual ISI plots.}
\label{figure_ISI}
\end{figure}

We consider a two-modulus 8-ary amplitude phase shift keying (APSK) signaling at \(30\) dB signal-to-noise-ratio (SNR), and use inter-symbol interference (ISI) metric for performance comparison averaged over 1000 channels randomly obtained using \texttt{sparse\_channel.m}. Equalizers are initialised such that the central tap is set to $1+\mathbf{i}$, and the rest of the taps are set to $(1+\mathbf{i})/N$ where \(N=120\) taps, and \(\mathbf{i}:=\sqrt{-1}\).
Results for ISI traces are summarised in Fig.~\ref{figure_ISI}, where the step-sizes appear in the legend. Note that the proposed equalizer RSCMA (with explicit $\ell_{1/2}$ closed form regularization) outperforms CMA and its sparse variants ANG-CMA and SCMA$(1/2)$ in terms of steady-state performance.

\section{Conclusions}

An $\ell_p$-regularised sparse CMA equalizer, RSCMA, is obtained and demonstrated for blind channel equalization of complex valued signals by incorporating the so-called zeroth-norm constraint in the traditional CM cost function.
Simulation results have shown that RSCMA exhibited faster convergence rate on sparse channels as compared to the traditional CMA equalizer and its sparse variants. Finally, our equalizer proved to be a substitute for the traditionally used ones.

\bibliographystyle{unsrt}
\bibliography{SparseCMA_Archiv}

\end{document}